\documentclass{PoS}

\usepackage{bm}
\usepackage{color}
\usepackage{array}
\usepackage{graphicx}
\usepackage{multirow}
\newcolumntype{P}[1]{>{\centering\arraybackslash}p{#1}}
\newcolumntype{M}[1]{>{\centering\arraybackslash}m{#1}}

\newcommand{\be}{\begin{equation}}
\newcommand{\ee}{\end{equation}}
\newcommand{\een}{\end{subequations}}
\newcommand{\ben}{\begin{subequations}}
\newcommand{\beq}{\begin{eqalignno}}
\newcommand{\eeq}{\end{eqalignno}}

\newcommand{\lsim}{\mathrel{\mathop{\kern 0pt \rlap
      {\raise.2ex\hbox{$<$}}}\lower.9ex\hbox{\kern-.190em $ \sim$}}}
\newcommand{\gsim}{\mathrel{\mathop{\kern 0pt
      \rlap{\raise.2ex\hbox{$>$}}}\lower.9ex\hbox{\kern-.190em $\sim$}}}

\title{Effective models of WIMP Direct Detection in DAMA/LIBRA--phase2}

\ShortTitle{Effective models of WIMP Direct Detection in DAMA/LIBRA--phase2}

\author{Sunghyun Kang, Stefano Scopel, Gaurav Tomar, \speaker{Jong-Hyun Yoon}\\
        Department of Physics, Sogang University, Seoul, Korea, 121-742\\
        E-mail:  \email{francis735@naver.com},  \email{scopel@sogang.ac.kr},  \email{tomar@sogang.ac.kr}, \email{jyoon@sogang.ac.kr}}

\abstract{Since the DAMA/LIBRA collaboration released updated
  results from their search for the annual modulation signal expected
  from Dark Matter (DM) scattering in their NaI detectors, we have
  fitted the updated DAMA result for the modulation amplitudes for a Weakly Interacting Massive Particle (WIMP) signal,
  parameterizing the interaction with nuclei in terms of the most
  general effective Lagrangian for a WIMP particle spin 1/2,
  systematically assuming dominance of one of the 14 possible
  interaction terms and a
  standard Maxwellian for the WIMP velocity distribution. We find that most of the couplings of the
  non--relativistic effective Hamiltonian can provide a better fit
  compared to the standard Spin Independent interaction case, and with
  a reduced fine--tuning of the three parameters (WIMP mass,
  WIMP--nucleon effective cross-section and ratio between the
  WIMP--neutron and the WIMP--proton couplings). In addition, effective
  models for which the cross section depends explicitly on the WIMP
  incoming velocity can provide a better fit of the DAMA data at large
  values of $m_{\chi}$ compared to the standard velocity--independent
  cross--section due to a different phase of the modulation
  amplitudes.  All the best fit solutions are in tension with
  exclusion plots of both XENON1T and PICO60.}

\FullConference{ICHEP 2018, XXXIX International Conference on High Energy Physics\\
		4-11 July 2018\\
		COEX, Seoul, Korea}

\begin{document}

\section{Introduction}
\label{sec:introduction}

In this work we will consider the most general WIMP (Weakly Interacting Massive Particle) --nucleus
effective Lagrangian for a WIMP particle of spin 1/2
scattering elastically off nuclei, systematically assuming the
dominance of one of the 14 possible interaction terms of the most
general non--relativistic Hamiltonian invariant by Galilean
transformations, fitting the new DAMA data to
the three parameters $m_{\chi}$ (WIMP mass), $\sigma_p$ (WIMP--nucleon
effective cross-section) and $c^n/c^p$ (WIMP--neutron/proton coupling ratio).

\section{WIMP rates in non--relativistic effective models}
\label{sec:eft}
Making use of the non--relativistic EFT approach one can write the most general Hamiltonian density
describing the WIMP--nucleus interaction as:

\begin{eqnarray}
{\bf\mathcal{H}}({\bf{r}})&=& \sum_{\tau=0,1} \sum_{j=1}^{15} c_j^{\tau} \mathcal{O}_{j}({\bf{r}}) \, t^{\tau} ,
\label{eq:H}
\end{eqnarray}
\noindent where $\mathcal{O}_{j}$ can be found in \cite{haxton1}.
Assuming that the nuclear interaction is the sum of the interactions
of the WIMPs with the individual nucleons in the nucleus the WIMP
scattering amplitude on the target nucleus $T$ can be written in the
compact form:

\begin{equation}
  \frac{1}{2 j_{\chi}+1} \frac{1}{2 j_{T}+1}|\mathcal{M}|^2=
  \frac{4\pi}{2 j_{T}+1} \sum_{\tau=0,1}\sum_{\tau^{\prime}=0,1}\sum_{k} R_k^{\tau\tau^{\prime}}\left [c^{\tau}_j,(v^{\perp}_T)^2,\frac{q^2}{m_N^2}\right ] W_{T k}^{\tau\tau^{\prime}}(y).
\label{eq:squared_amplitude}
\end{equation}

\noindent In the above expression $j_{\chi}$ and $j_{T}$ are the WIMP
and the target nucleus spins, respectively, $q=|\vec{q}|$ while the
$R_k^{\tau\tau^{\prime}}$'s are WIMP response functions which depend on the couplings $c^{\tau}_j$ as well as the transferred
momentum $\vec{q}$ and 
$(v^{\perp}_T)^2$. In equation (\ref{eq:squared_amplitude}) the
$W^{\tau\tau^{\prime}}_{T k}(y)$'s are nuclear response functions
and the index $k$ represents different effective nuclear operators,
which can be at most eight:
$k$=$M$,
$\Phi^{\prime\prime}$, $\Phi^{\prime\prime}M$,
$\tilde{\Phi}^{\prime}$, $\Sigma^{\prime\prime}$, $\Sigma^{\prime}$,
$\Delta$,$\Delta\Sigma^{\prime}$. The $W^{\tau\tau^{\prime}}_{T k}(y)$'s are function of $y\equiv (qb/2)^2$, where $b$ is the size of the nucleus. For the target nuclei $T$ used in
most direct detection experiments the functions
$W^{\tau\tau^{\prime}}_{T k}(y)$ have been calculated using nuclear shell models \cite{haxton1}. We systematically consider the possibility
that one of the couplings $c_{j}$ dominates in the effective
Hamiltonian of Eq. (\ref{eq:H}). In this case it is possible to
factorize a term $|c_j^p|^2$ from the squared amplitude of
Eq.(\ref{eq:squared_amplitude}) and express it in terms of the {\it
  effective} WIMP--proton cross section:

\begin{equation}
\sigma_p=(c_j^p)^2\frac{\mu_{\chi{\cal N}}^2}{\pi},
  \label{eq:conventional_sigma}
\end{equation}

\noindent (with $\mu_{\chi {\cal N} } $ the WIMP--nucleon reduced mass)
and the ratio $r\equiv c_j^n/c_j^p$.

\section{Analysis}
\label{sec:analysis}
\begin{figure}
 \includegraphics[width=0.5\columnwidth]{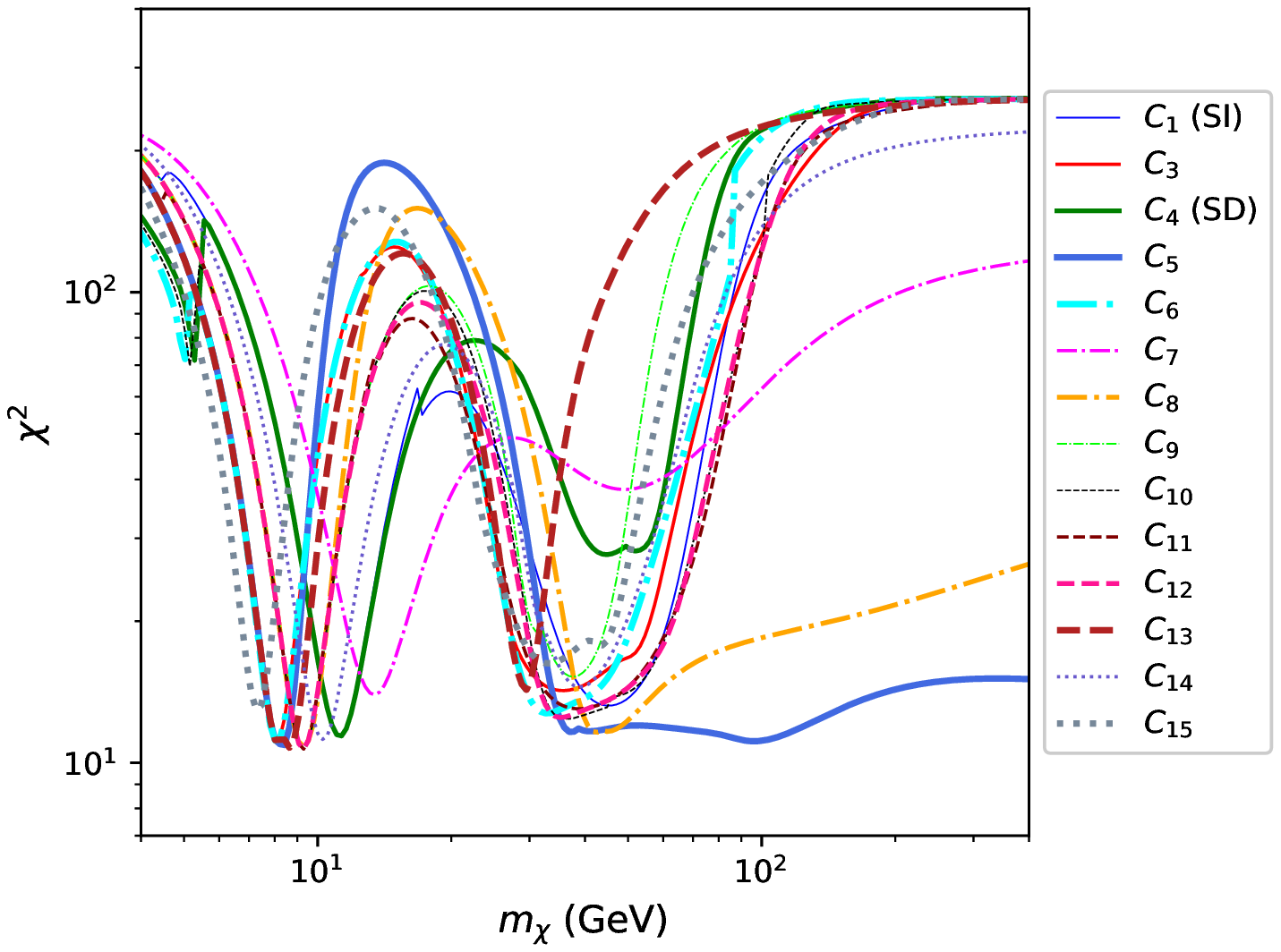}
 \includegraphics[width=0.5\columnwidth]{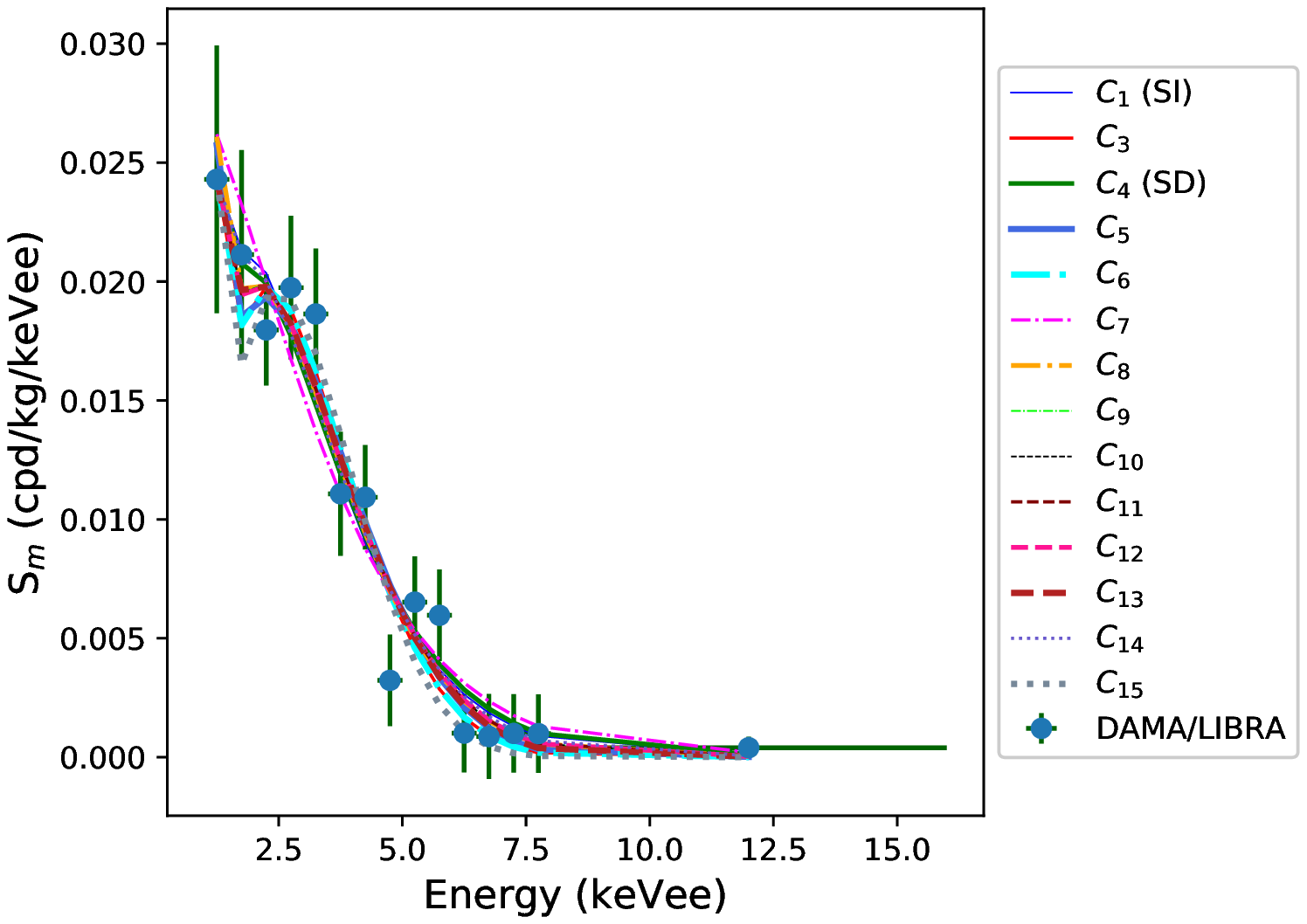}
\caption{Minimum of the $\chi^2$ of Eq.(\ref{eq:chi2}) at fixed WIMP
  mass $m_\chi$ as a function of $m_\chi$ for different WIMP-nucleus
  interactions (Left).
  DAMA modulation amplitudes as a function of the measured
  ionization energy $E_{ee}$ for the absolute minima of each effective
  model (Right). The points with error bars correspond to the combined data of
  DAMA/NaI, DAMA/LIBRA--phase1 and DAMA/LIBRA--phase2 \cite{dama_2018}}
\label{fig:sm_tot}
\end{figure}

\begin{figure}
\begin{center}
  \includegraphics[width=0.4\textwidth]{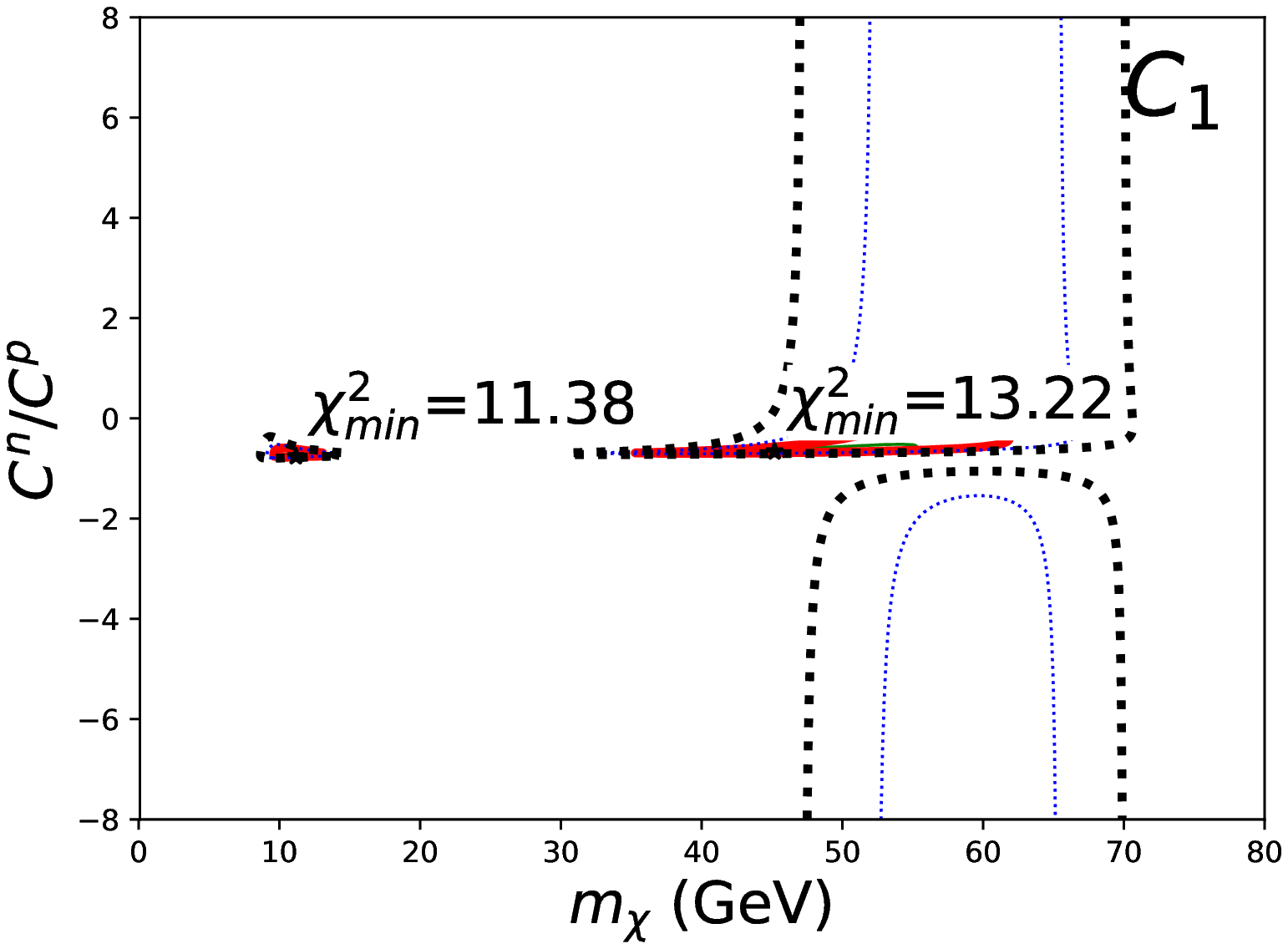}
 \includegraphics[width=0.4\textwidth]{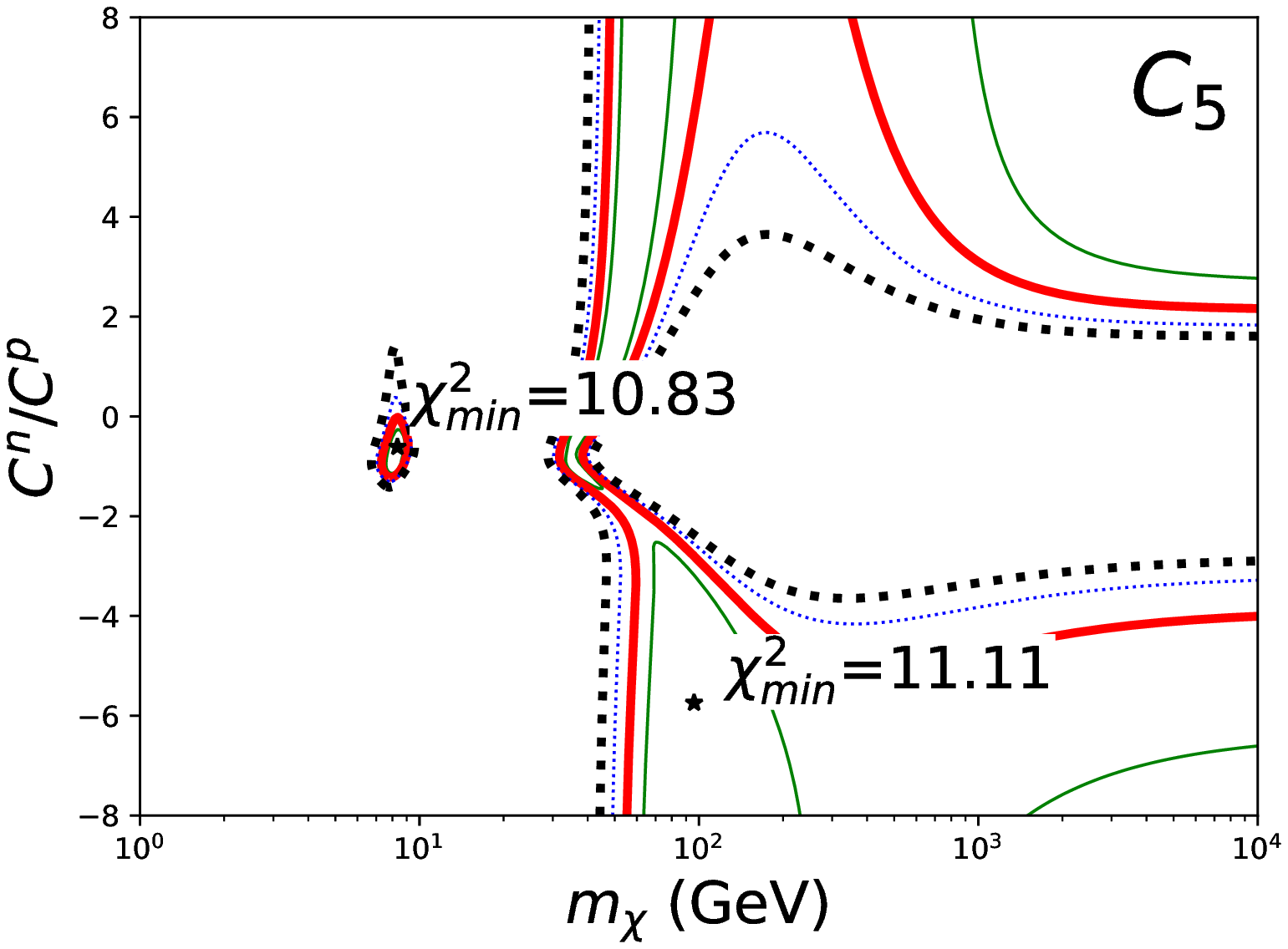}
 \end{center}
\caption{Contour plots of the $\chi^2$ of Eq.(\ref{eq:chi2}) minimized
  with respect to $\sigma_p$ in the $m_\chi$-$r$ plane for interactions $c_1$ and $c_5$. The thin (green) solid lines,
  the thick (red) solid lines, the thin (blue) dotted lines, and the
  thick (black) dotted lines correspond to 2, 3, 4, and 5 $\sigma$
  regions respectively. See \cite{dama_eft} for complete analysis.}
\label{fig:chi2_m_r_planes}
\end{figure}

We perform a $\chi^2$ analysis constructing the quantity:

\begin{equation}
\chi^2(m_{\chi},\sigma_p,r)=\sum_{k=1}^{15} \frac{\left [S_{m,k}-S^{exp}_{m,k}(m_{\chi},\sigma_p,r) \right ]^2}{\sigma_k^2}
  \label{eq:chi2}
  \end{equation}

\noindent and minimize it as a function of $(m_{\chi},\sigma_p,r)$. In the equation above $S_{m,k}$ and $\sigma_k$ represent the modulation amplitudes and errors measured by DAMA, while $S^{exp}_{m,k}$ denotes the expected modulation rate with respect to $(m_{\chi},\sigma_p,r)$.
In Fig. \ref{fig:sm_tot} we show the result of such minimization at
fixed WIMP mass $m_{\chi}$. The details of such
minima in the particular example of $c_1$ and $c_5 $ are provided in Table \ref{tab:best_fit_values}. The best fit parameters for $c_5$ appear to be less tuned compared
to the SI case. This can be seen in Fig.~\ref{fig:chi2_m_r_planes},
where we provide the contour
plots of the $\chi^2$ in the $m_{\chi}$--$r$ plane for $c_1$ and $c_5$. In particular, the regions within 2 and 3 $\sigma$ for
$c_1$ appear strongly tuned to the value $r$=-0.76, corresponding to
a cancellation in the WIMP-iodine cross section, whereas for $c_5$ the corresponding contour encompasses
a much wider volume of the parameter space.

\begin{table}[ht!]
\begin{center}
\begin{tabular}{|M{1cm}|M{3cm}|M{2cm}|M{2cm}|M{2cm}|} \hline
\rule[-10pt]{0pt}{20pt}  $\mathbf{c_j}$  & $\mathbf{m_{\chi,min}}$ \textbf{(GeV)} & $\mathbf{r_{\chi,min}}$ & $\mathbf{\sigma~(\mbox{\bf cm}^2)}$ & $\mathbf{\chi^2_{min}}$ \\\hline
 {$c_1$}     &  11.17 & -0.76 &  2.67e-38 & 11.38 \\ 
  {$c_5$}      &  8.34 & -0.61  &  1.62e-29 &  10.83\\  \hline
\end{tabular}
\caption{Absolute minima of the $\chi^2$ (see
  Eq.(\ref{eq:chi2})) for the couplings $c_1$ and $c_5$ of the effective
  Hamiltonian (\ref{eq:H}). See \cite{dama_eft} for a full set of contents in the table.}
\label{tab:best_fit_values}
\end{center}
\end{table}

In the case of $c_5$
the velocity--independent term of the cross section depends
on the $\Delta$ response function, which is proportional to the
nucleon angular momentum content of the nucleus, favoring elements
which have an unpaired nucleon in a non $s$--shell orbital. Both
iodine and sodium have this feature, implying in this case no
large hierarchy between the cross sections off the two nuclei, hence less fine--tuning. Namely,
numerically the isoscalar response function at vanishing momentum
transfer $W^{00}_{T\Delta}(q\rightarrow 0)$ for sodium is a factor
$\simeq$ 0.25 smaller compared to that for iodine. As can be seen from Fig. \ref{fig:sm_tot} for most models the $\chi^2$ shows a steep rise at large $m_{\chi}$. In these cases, the predicted modulation amplitude is
given by the cosine transform of the rate,
which is a function of the $v_{min}$ parameter only, and turns out to
be negative for $v_{min}\lsim$ 200 km/sec. This implies a bad fit
to the data and a large $\chi^2$.  The situation is different when the
cross section shows a non--negligible dependence on $v^\perp$,
(i.e. for models such as $c_5$). In this case the
integral of the differential rate is dominated by large values of $v>$
200 km/sec irrespective of $v_{min}$ and positive modulation
amplitudes can be obtained in the energy range of the DAMA signal also
at large values of $m_{\chi}$, implying in such regime a milder
increase of the $\chi^2$ (or even an acceptable fit in the specific
case of $c_5$).

\section{Conclusions}
\label{sec:conclusions}

The DAMA collaboration has released first results from the upgraded
DAMA/LIBRA-phase2 experiment \cite{dama_2018}. Assuming the dominance of one of the 14 possible
interaction terms of the Hamiltonian of Eq.(\ref{eq:H}), we have fitted the
experimental amplitudes to the three parameters $m_{\chi}$ (WIMP
mass), $\sigma_p$ (WIMP--nucleon effective
cross-section) and $c^n/c^p$ (neutron over proton coupling) with the WIMP velocity distribution a standard Maxwellian.
In Section \ref{sec:analysis}, we have discussed the phenomenological aspects of two particular effective operators, $c_1$ and $c_5$, in light of the results from DAMA/LIBRA--phase2.
In \cite{dama_eft}, we have studied all the other effective couplings of the Hamiltonian of Eq. (\ref{eq:H}) and found that all the best fit
solutions are in tension with exclusion plots of both XENON1T and
PICO60.


\begin{thebibliography}{99}

\bibitem{dama_2018}
R.~Bernabei et~al., {\it {First model independent results from
  DAMA/LIBRA-phase2}},  \href{http://arxiv.org/abs/1805.10486}{{\tt
  arXiv:1805.10486}}.

\bibitem{haxton1}
A.~L. Fitzpatrick, W.~Haxton, E.~Katz, N.~Lubbers, and Y.~Xu, {\it {The
  Effective Field Theory of Dark Matter Direct Detection}},  {\em JCAP} {\bf
  1302} (2013) 004, [\href{http://arxiv.org/abs/1203.3542}{{\tt
  arXiv:1203.3542}}].

\bibitem{dama_eft}
S.~Kang, S.~Scopel, G.~Tomar, and J.-H. Yoon, {\it {DAMA/LIBRA-phase2 in WIMP
  effective models}},  {\em JCAP} {\bf 1807} (2018), no.~07 016,
  [\href{http://arxiv.org/abs/1804.07528}{{\tt arXiv:1804.07528}}].
  


\end{thebibliography}
\end{document}